\documentclass[proof]{WileyASNA-v1}
\articletype{Article Type}%

\received{11 December 2022}
\revised{02 January 2023}
\accepted{10 January 2023}

\usepackage{natbib}
\usepackage{amsmath}

\newcommand\nobs{27436 }
\newcommand\boutbursts{3 }
\newcommand\boeclipses{5 }
\newcommand\neclipses{106 }
\newcommand\jdbegin{2459709.39673 }
\newcommand\jdend{2459904.43000 }
\newcommand\dataspan{195.03 }

\begin{document}
\title{Outburst Behaviour of the Dwarf Nova CG Draconis}

\author[1,2]{Maxim Usatov*}
\author[2]{Jeremy Shears}
\authormark{USATOV \& SHEARS}

\address[1]{\orgname{CBU Research Institute}, \orgaddress{J\'{a}chymova 27/4, Star\'{e} M\v{e}sto, 110 00 Prague 1, \country{Czech Republic}}}
\address[2]{\orgname{British Astronomical Association}, \orgaddress{PO Box 702, Tonbridge, TN9 9TX, \country{United Kingdom}}}

\corres{*7a Gainsborough Road, London, W4 1NJ, United Kingdom.
\email{maxim.usatov@bcsatellite.net}}

\abstract{
During the British Astronomical Association (BAA) 2022 campaign, \nobs photometric observations of the dwarf nova (DN) CG Draconis were made, with \neclipses eclipses recorded. This work summarizes the new data available and provides updated ephemeris and commentary on the observed eclipse profiles. The orbital period found is $P_{orb}=4^\textup{h} 31^\textup{m} 38^\textup{s} \pm 1^\textup{s}$. Two types of quasi-periodic outbursts are identified: normal outbursts, of $\Delta V \approx 1.25$ mag amplitude, and bright, of $\Delta V \approx 1.5$ mag. The pattern resembles superoutbursts of SU UMa-type DNe, however, no presence of superhumps characterizing these DNe was found. Given CG Dra is located above the period gap, it may represent a new intermediary subtype between SS Cyg and SU UMa-type stars, or provide support to superoutburst models that do not rely on eccentric accretion disks. 
}


\maketitle

\section{Introduction} \label{intro}
CG Dra has remained an enigmatic object for more than half a century. Initially found on the photographic plates taken by K. Loechel in 1964 with the 1.34-m Schmidt telescope of the Karl Schwarzschild Observatory, \citet{1966AN....289....1H, 1966AN....289..139H} lists CG Dra under preliminary designation S 9370 as U Gem-like, ``most likely belonging to the CN Ori group.'' U Gem stars are dwarf novae (DNe)---a particular type of cataclysmic variable (CV) consisting of a white dwarf (WD) primary star accreting matter from a red dwarf secondary. Comprehensive literature reviews of DNe are available from \citet{1995cvs..book.....W} and \citet{2001cvs..book.....H}. U Gem DNe are generally divided into four subtypes: SS Cyg, SU UMa, Z Cam and WZ Sge. While the latter two are completely inapplicable classifications in the case of CG Dra due to the lack of standstills and its long orbital period, CG Dra is usually referred to as a SS Cyg star \citep{1997A&A...325..601B, 10.1093/pasj/56.sp1.S1}. This type of DN is characterized by periodic outbursts, $\Delta V \sim$ 2--6 mag, in some cases with plateau, outburst rise time of a few d and a slightly longer decline, outburst interval of 10 d to a few yr, and orbital period, $P_{orb} > 3$ h. It is now widely accepted that DN outbursts occur due to the thermal limit cycle instability in the primary's accretion disk (AD), in particular due to the pile-up of material flowing from the secondary at a rate exceeding the rate of accretion in quiescence \citep{1974PASJ...26..429O, 1984AcA....34..161S}. At some point the disk becomes hot and ionized, this boosts the system's luminosity, increases disk viscosity, which then causes the orbiting material to spread, due to the exchange of angular momentum, partially inwards, falling towards the WD primary. The increased accretion rate quickly drains the AD and the system is returned to the cool, quiescent state.

The other subtype of DN, SU UMa, is characterized by the addition of periodic superoutbursts, typically brighter by $\sim$ 2 mag and occurring about $3\times$ less frequently than normal outbursts, and a shorter $P_{orb}<3$ h. All SU UMa-type DNe exhibit superhumps---periodic modulations appearing near superoutburst maximum with a period a few percent longer than $P_{orb}$. While normal SU UMa outbursts are believed to be caused by the same mechanism as in SS Cyg stars, multiple models exist explaining superoutbursts and superhumps: the thermal-tidal instability (TTI) model by \citet{1989PASJ...41.1005O, 1996PASP..108...39O}, the enhanced mass transfer (EMT) model currently supported by \citet{1991AcA....41..269S, 2004AcA....54..221S, 2017AcA....67..273S}, and the pure thermal instability (PTI) model by \citet{2010ApJ...725.1393C, 2012ApJ...747..117C}. 

Overviews of these models and description of associated problems are available from \citet{2013PASJ...65...50O} and \citet{2017AcA....67..273S}. In brief, the TTI model requires eccentric AD to explain superoutbursts. This condition occurs due to the 3:1 resonance in DNe with the mass ratio of components below a critical value, $q \equiv M_2/M_1 \lesssim q_{crit}$, where $q_{crit}\approx 0.3$, coinciding with the DN period gap between $P_{orb} \approx 2 \to 3$ h. \citet{2020AcA....70..313S} proposes $q_{crit} = 0.22$, which corresponds to the gap at $\approx 2.6$ h.  SU UMa DNe showing superoutbursts, hence, should not be found above the period gap. The EMT model explains superoutbursts as the result of variable hot spot brightness during enhanced mass transfer episodes; and the PTI model simulations suggest that thermal instability alone used to explain normal and long SS Cyg outbursts is sufficient to explain superoutbursts of SU UMa stars. At the current moment there is no consensus on which model is true. The TTI model has an issue of producing superhumps of excessive amplitude in simulations, and it is unable to explain them appearing in DNe above the period gap. 

CG Dra has entered BAA Variable Star Section (VSS) observing campaigns since 2001 as a poorly characterized DN. \citet{2007JBAA..117...22S} have identified its outburst period of $\approx$ 11 d and noted the bi-modality of outbursts. In the time-resolved photometry of \citet{2008JBAA..118..343S}, $P_{orb}=0.18864 \pm 0.00004$ d, noting the system's short and shallow grazing eclipses, consistent with its high inclination. The 14 eclipses recorded in both quiescent and outburst states appeared symmetrical, with both eclipse duration and depth independent of the state. The flickering appeared to be continuing throughout eclipses, suggesting that the system's inner AD is not occulted. In the earlier radial velocity measurements, \citet{1997A&A...325..601B} have found two principle power spectrum peaks corresponding to $P_{orb}=0.1893 \pm 0.0006$ and $0.2343 \pm 0.0021$ (d). 

In this work we present new photometric data, collected during a BAA VSS observing campaign in 2022. As a result, the number of observed eclipses has increased significantly. Most of the data were obtained by the authors on the 0.43-m \textit{A1} telescope of Alnitak Remote Observatories in Nerpio, Spain, and additional observations come from BAA observers acknowledged below. The long-term light curve, eclipse profile comparison in various states of the system, its outburst cycle and the analysis of outburst decline rates are provided and discussed in \S \ref{photometry}. CG Dra eclipse ephemerides, orbital period and spectrograms of frequencies in the $P_{orb}$ frequency domain are revealed in \S \ref{periods}. Results and other challenges present with this system are discussed in \S \ref{discussion} and this work is concluded in \S \ref{conclusion}.

\section{Photometry} \label{photometry}
\subsection{Light Curve and Eclipse Profiles}

\begin{figure*}
	\centering
	\includegraphics[width=7in]{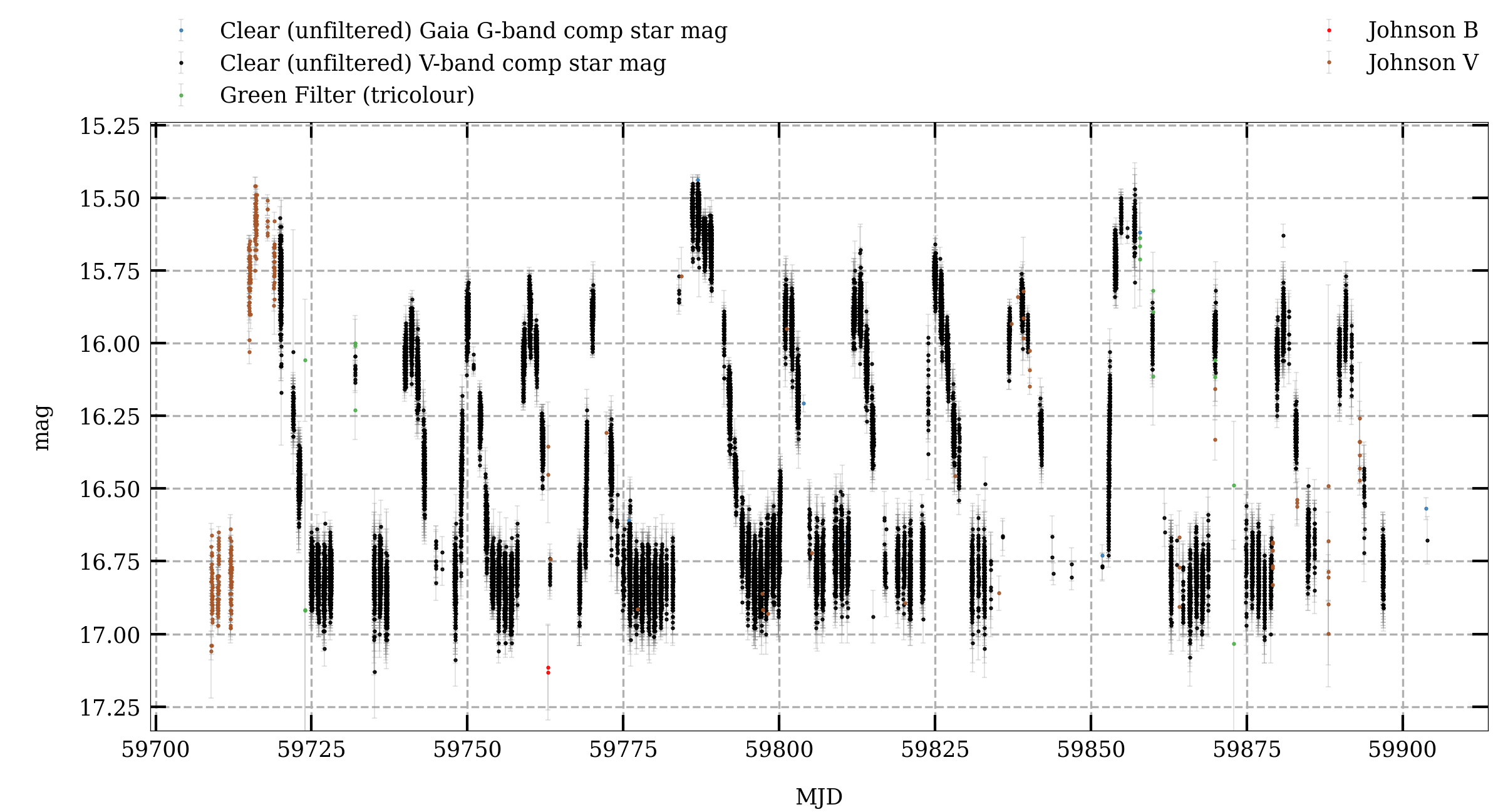} 
	\caption{The light curve of CG Dra obtained throughout the 2022 BAA VSS observing campaign. Filters are color-coded. The light curve spans JD 2459709.39673 to 2459904.43000 (195 d), containing \nobs observations. BAA observers who have contributed to the light curve above are D. G. Buczynski, D. Shepherd. F. Tabacco, G. Poyner, I. L. Walton, M. Mobberley, M. Usatov, N. D. James, P. Bouchier and R. Sargent.}
	\label{f:lightcurve}
\end{figure*}

CG Dra light curve obtained throughout the 2022 BAA VSS observing campaign is shown in figure \ref{f:lightcurve}. The data collected covers JDs \jdbegin $\to$ \jdend spanning \dataspan d. A total of \nobs observations were made, with \neclipses eclipses recorded. The 0.43 m \textit{A1} telescope used is a Corrected Dall-Kirkham (CDK) optical design equipped with the latest generation back-illuminated Sony IMX455 CMOS 24 $\times$ 36 mm sensor with 3.76 $\mu$m square pixels. All CG Dra observations were taken in the 2 × 2 binned mode with 0.53”/px scale that is suitable for typical 1-2” FWHM seeing conditions available at the site. Photometric reduction was done using \textsc{MetroPSF} Python code\footnote{The original \textsc{MetroPSF} source code is available at https://github.com/blackhaz/MetroPSF}, modified for batch processing. \textsc{SExtractor} routines \citep{1996A&AS..117..393B} were used for flux determination. \textsc{MetroPSF} performed automatic blind astrometric calibration via local copy of the Astrometry.net service \citep{2010AJ....139.1782L}, requested comparison photometry data from the AAVSO Photometric All-Sky Survey (APASS DR9) catalog \citep{2016yCat.2336....0H}, matched stellar sources and performed differential photometry of CG Dra via linear regression fits to a weighted V-band ensemble \citep{2010JAVSO..38..202P}. The ensemble was limited to known constant stars with AAVSO Unique Identifiers (AUIDs), within $\pm 3$ mag of CG Dra. All known variable stars in the AAVSO VSX database were excluded from the ensemble. The selection of stars was fixed throughout each night whenever possible. Typically 5-6 comparison stars were used per night, and one check star not part of the ensemble for control. Photometry with uncertainty exceeding $\pm 0.1$ mag was discarded, and good observations were submitted manually to BAA VSS and AAVSO databases after visual control. 

Most of observations were made without filters (CV mode) due to faint magnitudes in quiescence and short cadence requirement. Only a short period of observations of the first bright outburst was taken with the Johnson V filter. All exposures were of 30 s duration. 
CG Dra is quiescent near 17 mag, bursting to $\approx 15.75$ mag. Visually two outburst types can be identified---normal and bright, the latter being $\approx 0.25$ mag brighter. We have captured \boutbursts bright outbursts, and the outburst cycle is analyzed in \S \ref{outburstcycle}. 

\begin{figure}
\centering
\includegraphics[width=3.4in]{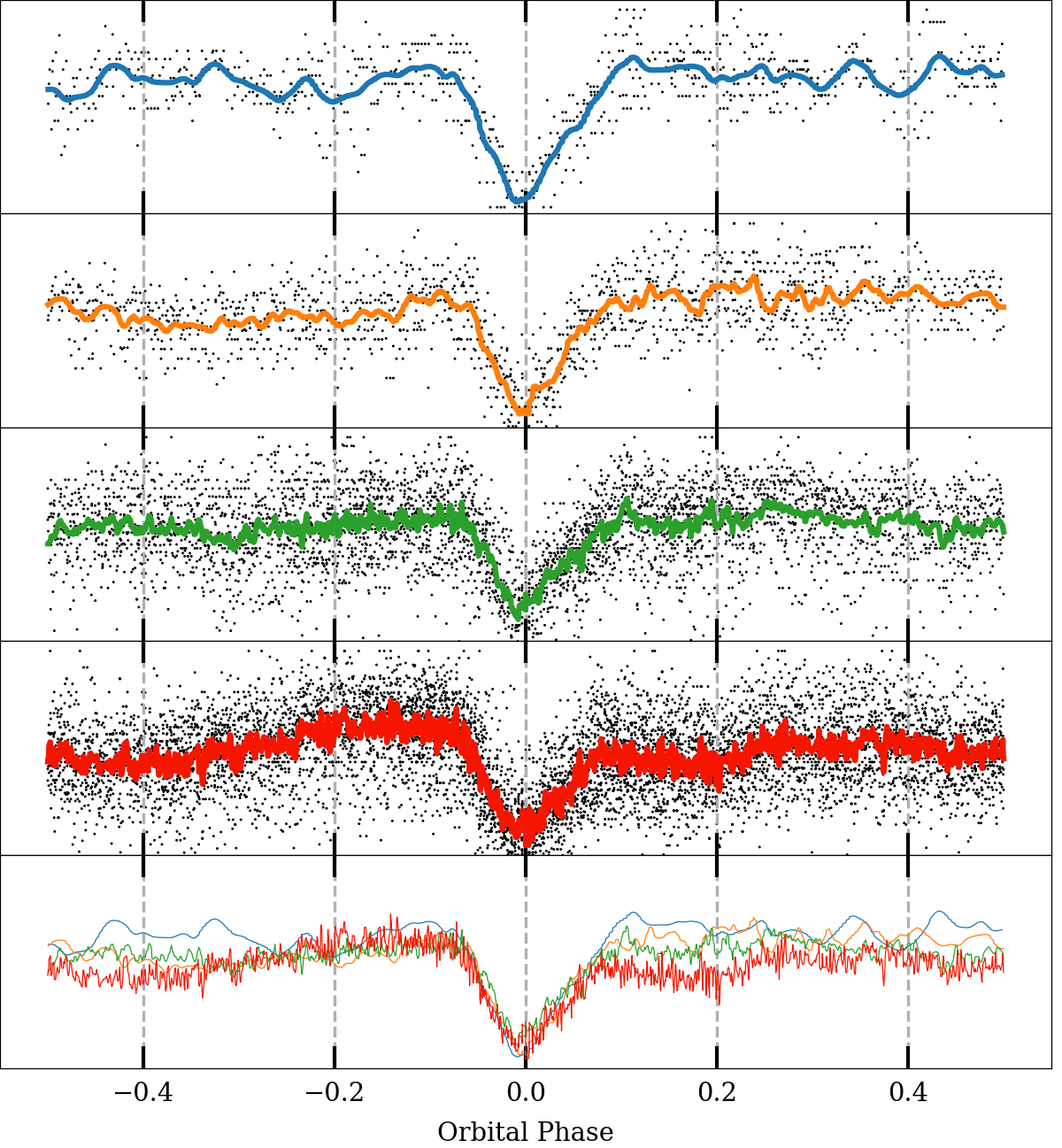} 
\caption{Phase plot of CG Dra normalized average eclipse profiles, in different system states. From top to bottom: bright outbursts, normal outbursts, rising and fading, and quiescence. Ordinates are in the [0, 1] range. Black points represent individual observations of \neclipses eclipses. The bottom pane compares averaged eclipse profiles.}
\label{f:eclipseprofiles}
\end{figure}
Averaged and normalized profiles of eclipses are shown in figure \ref{f:eclipseprofiles}. All eclipses were divided into bright outburst, outburst, rising and falling, and quiescent states manually. The system was considered in outburst state if $V \lesssim 16.25$ and in bright outburst state if $V \lesssim 15.75$. Eclipses recorded on the decline from these states were considered fading. Quiescent state assumed once the system reached $V \approx 16.50$, and rising once $V \lesssim 16.50$ or if the linear slope of a nightly light curve showed significant negative linear trend. Ephemerides found in \S \ref{periods} were used to calculate predicted eclipse minima. Each eclipse light curve was truncated to the length of orbital period, centered on its calculated minimum, and smoothed using Gaussian filter with $\sigma = 0.2$. Here, the filter was applied in time domain on unevenly sampled data, and the standard deviation of the Gaussian distribution was found empirically to remove light curve flickering and other noise. Approximate eclipse boundaries were found around the dimmest magnitude point of the truncated smoothed light curve using a simple condition that intensity of a smoothed light curve can only increase with distance from the eclipse minimum towards each boundary. The light curve was truncated further to found boundaries, and a Gaussian function was fitted to the remaining data. The minimum of the fit was used to record the observed eclipse minimum. Eclipses were folded using $P_{orb}$ found in \S \ref{periods} and the observed minimum epoch. For each state of the system, folded light curves were combined into unevenly sampled data sets which were averaged using Gaussian filter with $\sigma = 5$, found empirically to produce a smoothly varying average eclipse profile per state. This provides a case of weighted averaging whereby data sets with fewer data points, i.e. higher inter-sampling time difference, receive a higher degree of smoothing.        

CG Dra exhibits a variety of eclipse profiles in line with canonical DN outburst models. Quiescent states are dominated by asymmetric eclipses with a lower egress flux, compared to ingress, and an orbital hump of $\Delta V \sim 0.1$ mag, extending throughout orbital phases $\phi \approx -0.4 \to 0.2$, representing the bright spot where the flow of material from the secondary impacts the primary's AD. Occasionally orbital hump amplitudes exceed 0.2 mag. The eclipse ingress appears to occur at $\phi \approx -0.07$ when the secondary begins occulting the AD and the bright spot. As noted by \citet{2008JBAA..118..343S}, the short duration of eclipses suggests that CG Dra's inclination is close to critical, hence its eclipses are grazing---occulting only the bright spot and, partially, the AD. The presence of flickering, typically of $\Delta V \sim 0.05$ mag, throughout the eclipses indicates that its likely source, the inner part of the AD, is not occulted. Quite often the flickering reaches $\approx 0.1$ and, sometimes, $0.2$ mag amplitudes. The egress occurs at $\phi \approx 0.08$. 

The rising and fading states occurring before and after normal outbursts are dominated by eclipses with low-amplitude ($\Delta V \lesssim 0.1$ mag) orbital hump and symmetric profiles, with a slightly gentler egress slope. The subsidence of orbital hump indicates that the hot and ionized AD contributes a higher proportion of the total system's light in these states, dominating their profiles. Symmetric low-amplitude hump eclipses are the most common throughout non-quiescent states. In bright outburst state the egress flux is higher than on the ingress. This is explained by the increasing presence of post-egress ``flares'' during this state, which could be attributed to flickering. 

\begin{figure}
\centering
\includegraphics[width=3.4in]{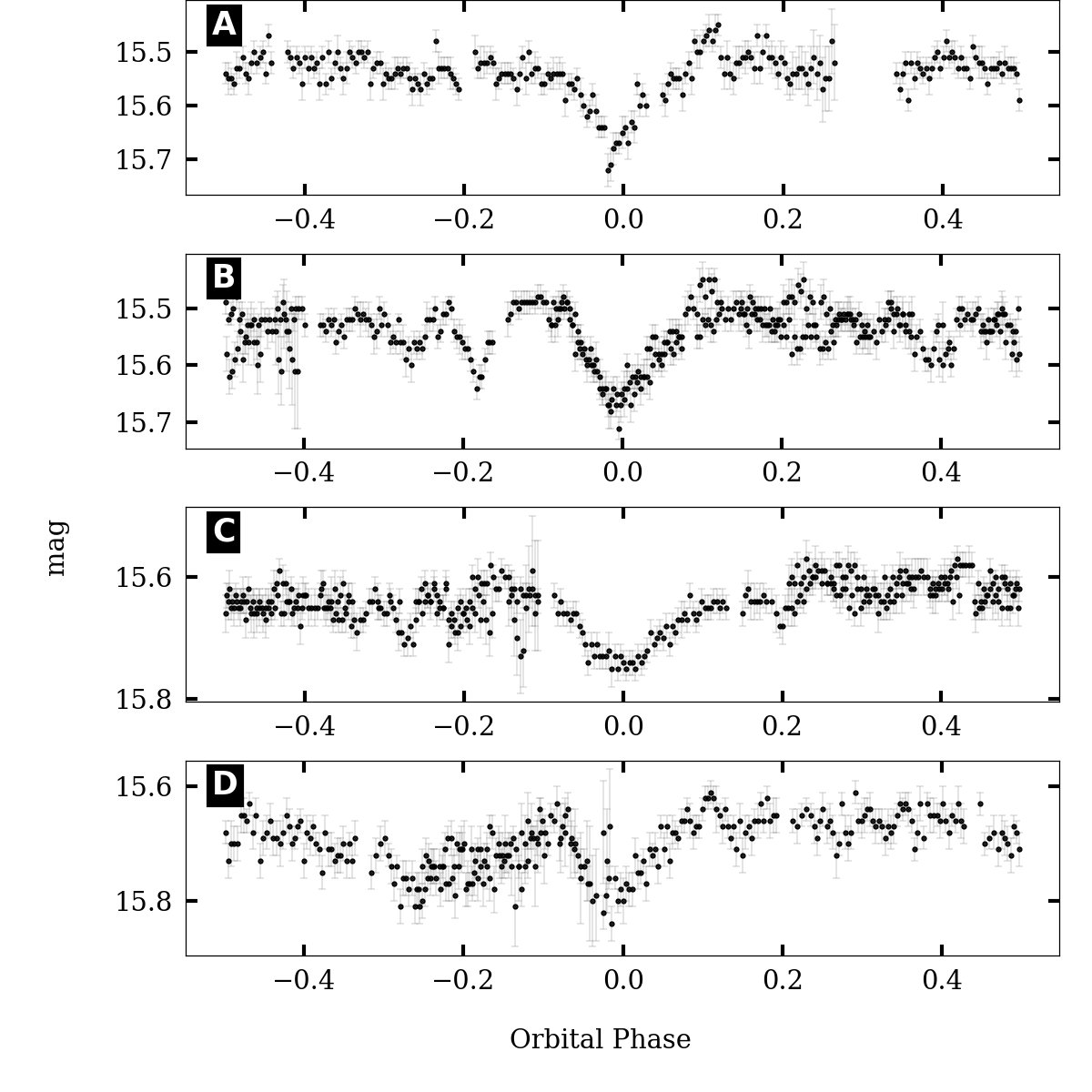} 
\caption{CG Dra light curves showing eclipses in its bright outburst state, with post-egress ``flares'' (panes A, B and D) and broad humps (panes C and D).}
\label{f:brightoutbursts}
\end{figure}

\begin{figure}
\centering
\includegraphics[width=3.4in]{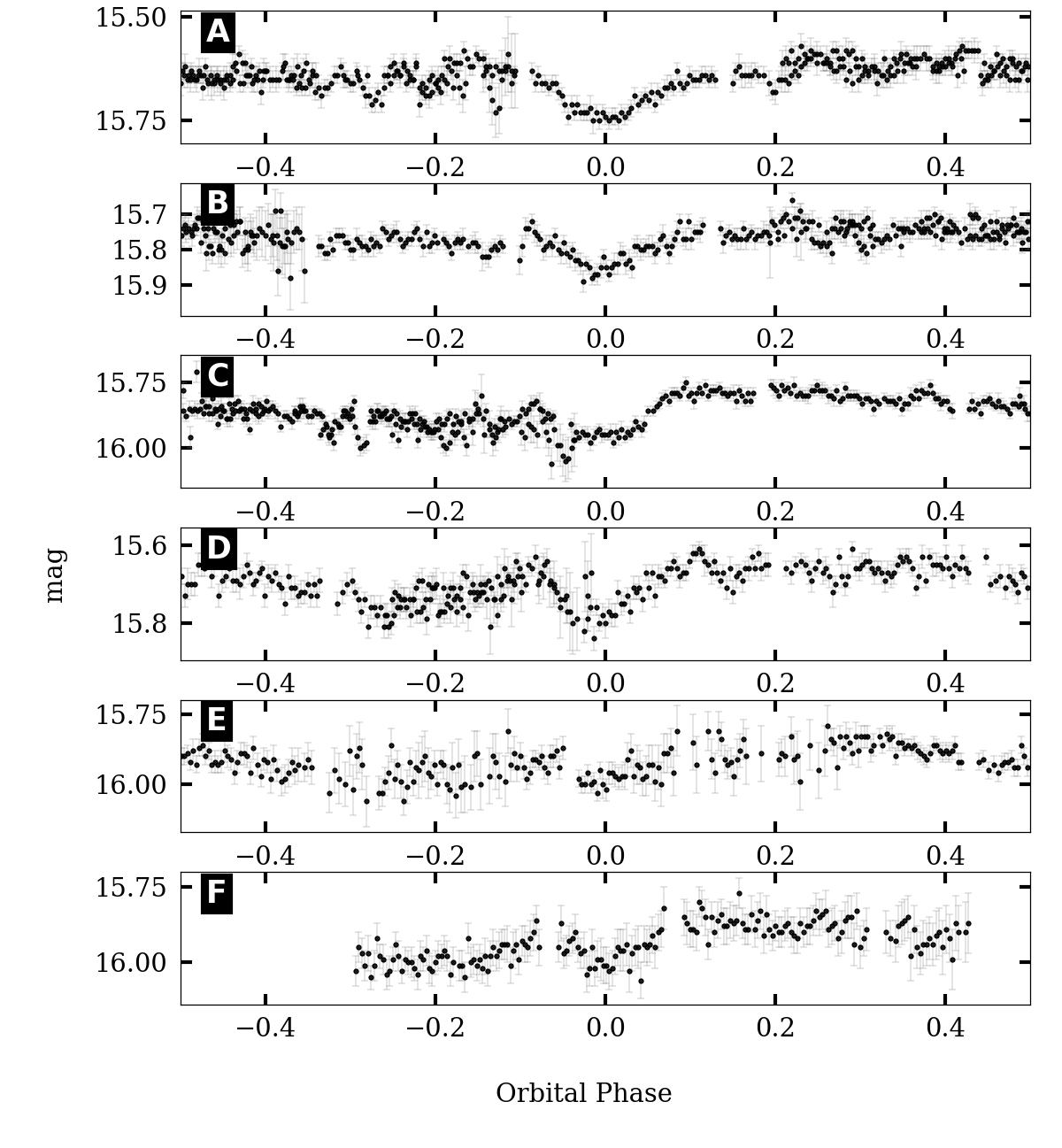} 
\caption{Occasionally CG Dra light curves show post-egress ``flares'' and broad post-egress humps. Phase plot, panes A and D: bright outburst, pane C: fading after normal outburst, panes B, E and F: normal outburst. }
\label{f:pehs}
\end{figure}

\begin{figure}
\centering
\includegraphics[width=3.4in]{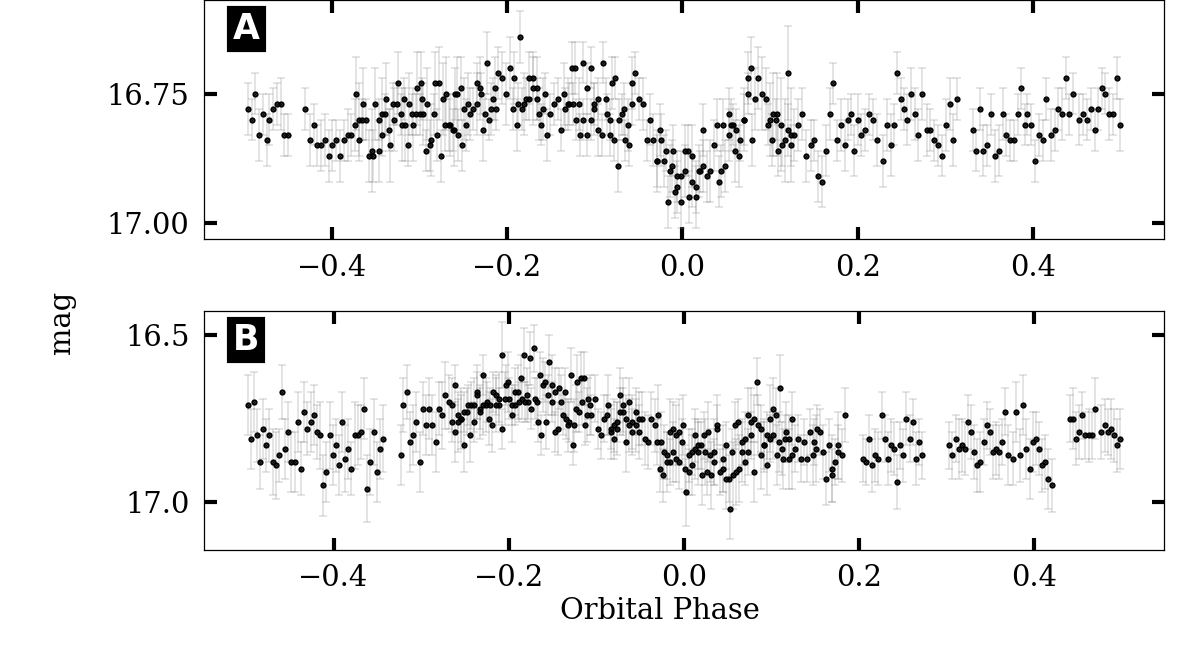} 
\caption{Highly asymmetric CG Dra eclipse profiles observed during fading to quiescent state, with egress magnitudes close or near the eclipse minima.}
\label{f:ha}
\end{figure}

CG Dra demonstrates a wide variety of eclipse profiles throughout all of its states. The new data shows that CG Dra ingress and egress phases vary from quiescence to bright outbursts, getting slightly wider as the system progresses from quiescence to the bright outburst state. This suggests that the size of the AD varies with the state of the system. Eclipse depth also increases in this sequence. Both effects are more pronounced at the point of egress. Although only \boeclipses eclipses have been recorded in the bright outburst state, their profiles are of particular interest, as they represent the system in its extreme state. These eclipses are shown in figure \ref{f:brightoutbursts}. Data cadence is 30 s for all figures mentioned below. Common in this state are post-egress ``flares'' occurring immediately after the egress---see panes A, B and D for example, and broad post-egress humps, shown in panes C and D. Post-egress humps and ``flares'' occasionally appear not only during the bright outburst state---for example see pane C in figure \ref{f:pehs} for those in fading state after a normal outburst, and, also, panes B, E and F showing broad post-egress humps during normal outbursts. High-amplitude ``flares'' can be explained by the enhanced accretion flow and, hence, strong flickering in this state. However, the nature of post-egress humps at $\phi \approx 0.1 \to 0.5$ is uncertain. In contrast to eclipses with post-egress humps are highly asymmetric eclipses with egress magnitudes only slightly above the preceding eclipse minima---see examples in figure \ref{f:ha}. Four such eclipses were observed in the fading state of CG Dra, all on the final return to the quiescent state. 

\subsection{Outburst Cycle and Decline Rate} \label{outburstcycle}

\begin{figure}
\centering
\includegraphics[width=3.4in]{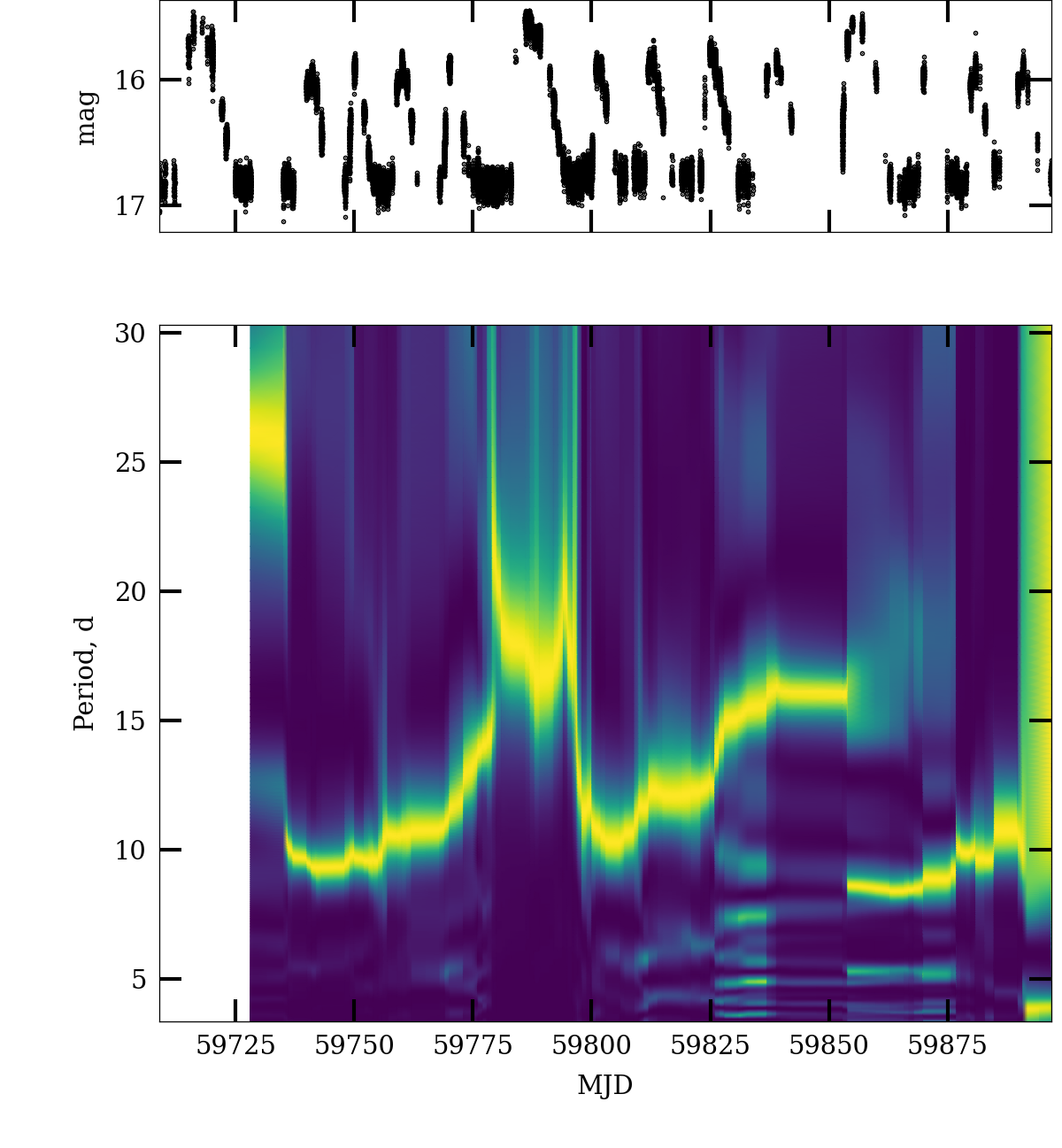} 
\caption{Two-dimensional AoV spectrogram of frequencies in the domain of the CG Dra outburst cycle. Normal outbursts occur every $\sim$ 10 d.}
\label{f:os}
\end{figure}

CG Dra follows a quasiperiodic outburst cycle. A two-dimensional spectrogram of frequencies in the domain of the outburst cycle is shown in figure \ref{f:os}. The method to compute the spectrogram is similar to that described by \citet{2013PASJ...65...50O}, except for a few changes. Non-detrended normalized data is used as the input, with spectrogram step equal to $1/300$ and the window of $1/6$ of the whole dataset, without smoothing. An orthogonal multi-harmonic analysis of variance (AoV) algorithm was used to find periods \citep{1996ApJ...460L.107S} implemented in the \textsc{P4J} Python module \citep{2018ApJS..236...12H}.An AoV periodogram was computed for each step with resolution $R=1000$, and the number of harmonics, $n_h = 1$. Some discontinuities are visible along the time axis of the spectrogram which are attributed to the uneven sampling of the data. 

\citet{2001IBVS.5124....1K}, observing two CG Dra outbursts, noted that CG Dra has an unusual outburst decline rate for its orbital period---0.14 and 0.31 mag d$^{-1}$. DNe are known to follow a well-defined decay time--$P_{orb}$ relation \citep{1975JBAA...86...30B, 1995cvs..book.....W}:

\begin{equation}
\tau_d = 0.53 P_{orb}^{0.84} \textup{(h) d mag}^{-1}.
\end{equation}

For CG Dra, applying $P_{orb}$ found in \S \ref{periods}, the expected decay time scale $\tau_d = 1.88$ d mag$^{-1}$, corresponding to 0.53 mag d$^{-1}$ expected decline rate. The derivative of CG Dra magnitude resampled, smoothed and interpolated to 1-hour bins indicates decline rates peak at $\sim 0.36$ and average at $\approx 0.25$ (mag d$^{-1}$), approximately half way to quiescence. Rise rates are about twice as fast. This confirms that CG Dra deviates from the expected relation for DNe.

\section{Ephemerides} \label{periods}
\subsection{Orbital Period and the O--C Chart}

In order to find the orbital period, the long-term CG Dra light curve was detrended using a locally weighted polynomial regression algorithm \citep{doi:10.1080/01621459.1979.10481038} implemented via \textsc{LOWESS} Python package by A. Lee. The bandwidth parameter $b=0.005$ was used which produces a smoothed light curve while preserving signals in the orbital period domain. The smoothed curve was subtracted from the original data to produce detrended light curve. The \textsc{P4J} AoV algorithm implementation was used to find the orbital period from the detrended data, with $R=10^4$ and $n_h=8$. The most prominent peak was refined with $R=10^5$. To estimate period error, as no interfering frequencies were detected and the level of the orbital period is quite strong, $S/N=76$, we adopt a simple procedure described by \citet{1991MNRAS.253..198S}, applicable to strong signals. The mean noise power level, $N^2$, was defined as the median of the spectrum in the vicinity of the frequency $f$ of the most prominent spectral line of power $p$, with boundaries set to $f\pm10\%$. The width of the spectral line at the $p-N^2$ level represents the $1\sigma$ confidence interval of the spectral line. The $P_{orb}$ found is 0.188640 $\pm$ 0.000007 d. This corresponds to $4^\textup{h} 31^\textup{m} 38^\textup{s} \pm 1^\textup{s}$. 

\begin{figure}
\centering
\includegraphics[width=3.4in]{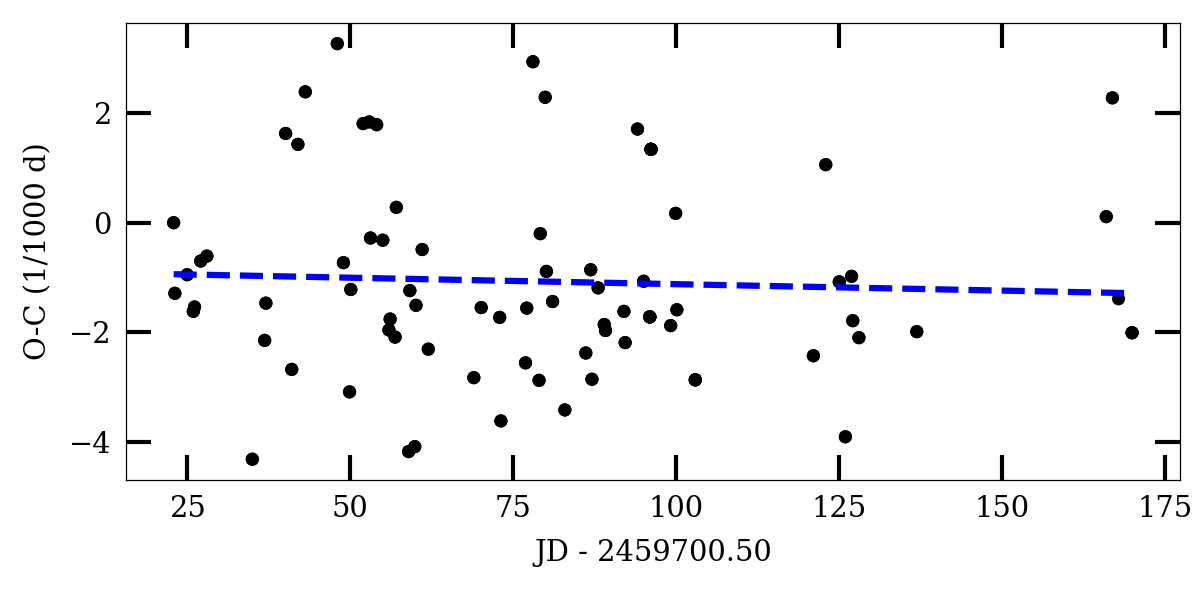} 
\caption{Observed minus calculated chart for CG Dra $P_{orb}=0.188640$ $\pm$ 0.000007 d. Fitted linear trend line is shown in blue.}
\label{f:oc}
\end{figure}

The $P_{orb}$ value was tested by constructing the observed minus calculated (O--C) chart, shown in figure \ref{f:oc}, comparing observed eclipse minima calculated as described in \S \ref{photometry} with predicted values. Eclipses with poor Gaussian fits were removed. No significant period change from test ephemeris was detected.    

To determine the epoch of minimum, the method of \citet{1956BAN....12..327K} was used, implemented in the \textsc{Peranso} software. High-quality light curve was selected of an eclipse in bright outburst state of CG Dra, with suppressed flickering and symmetric eclipse profile dominated by AD light. The following ephemeris was obtained: 

\begin{equation}
\textup{JD}_{min} = 2459788.508796(694) + E \times 0.188640(7).
\end{equation}

\subsection{Spectrogram in the Orbital Period Domain}
\begin{figure}
\centering
\includegraphics[width=3.4in]{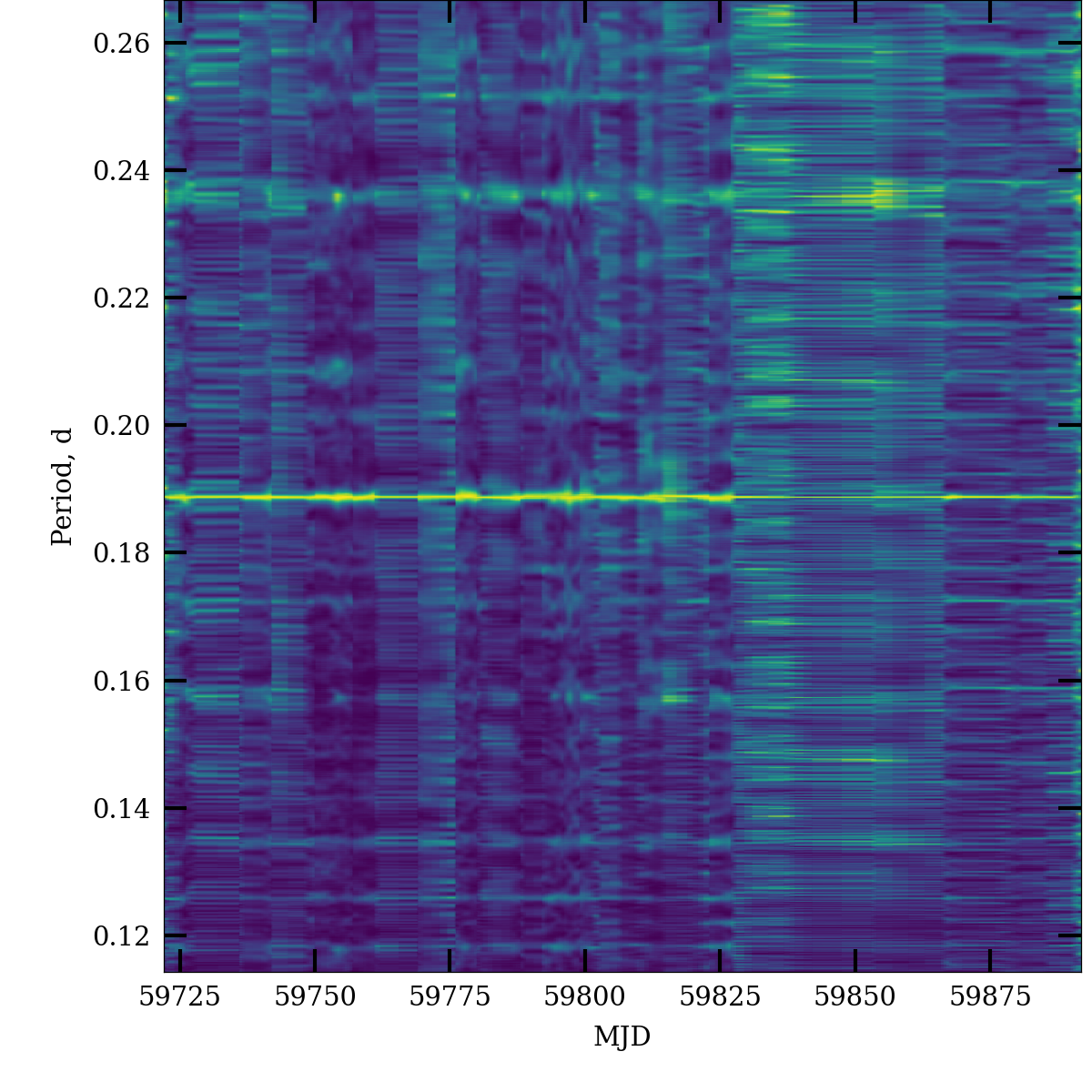} 
\caption{Two-dimensional AoV spectrogram of frequencies in the domain of the CG Dra orbital period, $P_{orb}=0.18864$ d, prominent in this chart.}
\label{f:porbspec}
\end{figure}

To search for potential superhump signatures, typically expected within a few percent of $P_{orb}$, we build a spectrogram within this domain of frequencies, shown in figure \ref{f:porbspec}. LOWESS-detrended data is used as the input, with spectrogram step equal to $1/100$ and the window of $1/20$ of the whole dataset, without smoothening. AoV parameters used were $R=1000$ and $n_h=8$. The orbital period is prominent in the spectrogram, along with a spectral line at $\approx$ 0.237 d, and a weaker one at $\approx 0.159$ d. These additional periodicities are likely associated with most data being taken at night at the same location, as they correspond to the true orbital frequency $f_{orb}=5.30 \pm 1$ (c d$^{-1})$. In other words, these lines likely represent an additional artificial diurnal cycle on both sides of the true frequency and are the aliases of the orbital frequency. 

As an additional test we constructed a phase-dispersion minimization (PDM) \citep{1978ApJ...224..953S} spectrogram of the same detrended data. The PDM algorithm used was implemented in the \textsc{Astrobase} Python package \citep{wbhatti_astrobase}. A spectrogram window of 1/20 and step of 1/40, of the whole data set were used, with $R=100$. Again, only the orbital period is prominent, with hints of artificial aliases mentioned above. No other significant periods are detected in the orbital period domain via both AoV and PDM methods.

\section{Discussion} \label{discussion}

CG Dra appears to exhibit two distinct types of outbursts, which we can call normal and bright outbursts. Bright outbursts resemble SU UMa-type superoutbursts, however, CG Dra does not show any superhump signatures and is located above the period gap, which makes its unambiguous classification difficult. No SU UMa-type stars have been found that show superoutbursts without superhumps. If bright outbursts are the equivalent of SU UMa superoutbursts then our observations might support models that do not specifically require eccentric ADs, or suggest that CG Dra may possibly be an intermediary U Gem subtype between SS Cyg and SU UMa.

Assuming CG Dra bright outbursts correspond to SU UMa superoutbursts, the lack of superhumps provides a challenge for the TTI model that requires superhumps to be present due to the eccentric AD. If superoutbursts appear due to the tidal dissipation in the eccentric disk then this also creates a problem, as CG Dra is located well above the period gap where these effects are not supposed to happen. Assuming CG Dra is related to the SS Cyg subtype, what then explains the appearance of bright outbursts? Long outbursts can actually occur in SS Cyg DNe---for example, the 45-day long outburst of U Gem in 1985, somewhat brighter than the normal outbursts of this star. \citet{2004AcA....54..433S} have, in fact, detected 0.3 mag amplitude superhumps in the 1985 data, ``appearing not later than 2--3 days after reaching maximum, and disappearing 4 days before final decline.'' This poses a challenge to the TTI model that is unable to explain superoutbursts and superhumps in systems with $P_{orb}$ above the period gap (U Gem $P_{orb} = 4.25$ h.)

\citet{2000A&A...353..244H} estimate that the mass accreted during the 1985 U Gem outburst was higher than the mass of the whole AD in quiescence. A likely trigger for such an outburst would have been the increased mass transfer rate from the secondary, probably caused by the irradiation of the secondary, supporting the EMT superoutburst model. However, this long outburst was an isolated episode, unlike the periodic bright outbursts we observe in CG Dra. 

SS Cyg itself exhibits bi-modal outburst distribution, with long outbursts lasting $>12$ d, and short, lasting $< 12$ d \citep{2007PASP..119.1361P}. These are generally sequenced LS (long-short) and, less commonly, LLS, LSSS or LLSS. The amplitude of long and short outbursts tend be the same, while there are also occasional anomalous short outbursts of smaller amplitude. To compare this with CG Dra, its normal and bright outbursts differ significantly in amplitude and, with two bright outburst cycles observed, 4--5 normal outbursts are seen between the bright ones.

There is an example of a DN which exhibits properties of different subtypes at the same time---NY Serpentis that shows three distinctive type of outbursts: normal, wide outbursts without superhumps, and superoutbursts with superhumps \citep{2014PASJ...66..111P}. NY Ser could be an intermediary DN subtype---a case potentially applicable to CG Dra.   

In addition to the problems presented above, CG Dra provides another challenge: unexpected spectral type of the secondary star. Although this work is focused on photometry, it is worth mentioning this problem, at least in brief. The first spectrophotometric observations of CG Dra were done by \citet{1990AGAb....4...24S, 1992A&A...266..225S} using the 2.2-m telescope at Calar Alto Observatory, and then later on 4.2-m William Herschel Telescope at La Palma by \citet{1997MNRAS.287..271S}, who identified the secondary's spectral type within the range of K5--7. \citet{1997A&A...325..601B}, observing with the 3.5-m telescope at Calar Alto, identified it as K5 $\pm$ 2. Assuming that the canonical CV mass-period relationship holds for CG Dra, the secondary mass can be estimated as 

\begin{equation}
M_2 = 0.065 P_{orb}^{5/4} (\textup{h})
\end{equation}

for $1.3 \leq P_{orb} (\textup{h}) \leq 9$ \citep{1984ApJS...54..443P}. For CG Dra, $M_2 \approx 0.43 M_{\odot}$. A main sequence star of this mass is expected to be of M3--M4 spectral type \citep{2018MNRAS.479.5491E}. A K5 secondary would be significantly overweight, requiring longer $P_{orb} \sim 7$ h.

Yet another spectroscopic issue with CG Dra is that its emission and absorption spectrum lines move in phase \citep{1997A&A...325..601B}. In CVs, the absorption component is generally attributed to the secondary star, and emission---to the hotter AD, thus an opposite is expected. No reasonable explanation of this observation exists. 

\section{Conclusion} \label{conclusion}

We have presented the most extensive photometric data set available on CG Dra to date. The orbital period measured was $P_{orb}=4^\textup{h} 31^\textup{m} 38^\textup{s} \pm 1^\textup{s}$. Throughout different states of the system a variety of eclipse profiles was observed, consistent with standard DN models with the exception of post-egress ``flares'' and broad humps. The presence of two types of quasi-periodic outbursts---normal and bright---was evident in the data. No superhump signatures were found in AoV and PDM spectrograms at frequencies in the domain of $P_{orb}$. Assuming the bright outbursts of CG Dra correspond to SU UMa-type superoutbursts, this creates an issue of explaining, within the TTI model, how superoutbursts appear without superhumps in a DN that is significantly above the period gap, where SU UMa-type stars should not be found. For both normal and bright outbursts, we confirm that the decline rates observed are too slow for known DN decay time--$P_{orb}$ relations. 

Interpreting CG Dra within canonical DN classification and models is problematic. Do post-egress ``flares'' during bright outburst and, occasionally and to a lesser extent, in normal outburst state, represent EMT episodes that lead to superoutbursts? What triggers bright outbursts in this system and are post-egress humps related to them? Further observations are required to understand the physical nature of this system. We hope that observations presented herein will be helpful for further research. The photometry obtained is openly available via BAA and AAVSO variable star databases.

\section*{Acknowledgements}
This research has made use of National Aeronautics and Space Administration (NASA) Astrophysics Data System Bibliographic Services and the SIMBAD database, operated at CDS, Strasbourg, France. The authors acknowledge with thanks the variable star observations from the AAVSO International Database and BAA Variable Star Section contributed by observers worldwide and used in this research.

\bibliography{Master}%

\end{document}